# A Compressive Sensing Approach for Connected Vehicle Data Capture and Recovery and its Impact on Travel Time Estimation

Lei Lin, Weizi Li, and Srinivas Peeta

*Abstract*— Connected vehicles (CVs) can capture and transmit detailed data such as vehicle position and speed through vehicle-to-vehicle and vehicle-to-infrastructure communications. The wealth of CV data provides new opportunities to improve safety and mobility of transportation systems. However, it is likely to overburden storage and communication systems. To mitigate this issue, we propose a compressive sensing (CS) approach that allows CVs to capture and compress data in real-time and later recover the original data accurately and efficiently. The approach is evaluated using two case studies. In the first study, we use this approach to recapture 10 million CV Basic Safety Message (BSM) speed samples. It can recover the original speed data with root-mean-squared error as low as 0.05. We also explore recovery performance for other BSM variables. In the second study, a freeway traffic simulation model is built to evaluate the impact of this approach on travel time estimation. Multiple scenarios with various CV market penetration rates, On-board Unit (OBU) capacities, compression ratios, arrival rate patterns, and data capture rates are simulated. The results show that the approach provides more accurate estimation than conventional data collection methods, through up to 65% relative reduction in travel time estimation error. Even when the compression ratio is low, the approach can provide accurate estimation, thereby reducing OBU hardware costs. Further, it can improve accuracy of travel time estimation when CVs are in traffic congestion as it provides a broader spatial-temporal coverage of traffic conditions and can accurately and efficiently recover the original CV data.

## I. INTRODUCTION

RECENT technological advances and their implications for socioeconomic benefits associated with improved traffic conditions have prompted widespread focus on connected vehicles (CVs). CVs have the potential to improve safety and mobility of the transportation system by enhancing situational awareness [1], [2] and traffic state estimation [3] through vehicle-to-vehicle (V2V) and vehicle-to-infrastructure (V2I) communications.

As an example of such endeavors, in 2012, the Safety Pilot Model Deployment (SPMD) program was launched in Ann Arbor, Michigan, United States. Nearly 3000 vehicles were equipped with GPS antennas and DSRC (Dedicated Short-Range Communications) devices. Each vehicle broadcasted Basic Safety Messages (BSMs), which included the position, velocity, and yaw rate, to nearby CVs and roadside units at a rate of 10 Hz [4]. This CV data provides opportunities for improving intelligent transportation systems based applications such as traffic state estimation and traffic signal optimization. However, the high sampling rate (i.e., 10 Hz), which can result in 25GB data being captured and uploaded every hour [5], leads to prohibitive storage and communications costs. According to Muckell et al. [6], the annual cost of tracking a fleet of 4000 vehicles would range from $1.8 million to $2.5 million. Due to the rapid increase in CV production and its increasing market penetration rate (MPR), this cost is expected to grow substantially in the near future.

To address the aforementioned challenge, previous studies have mainly taken two approaches. The first is called sample-then-compression, which collects data at a fixed rate in real-time and compresses the data offline. The Douglas-Peucker algorithm is one of the classical sample-then-compression methods [7]. Richter et al. (2012) introduced a semantic trajectory compression method, which utilizes reference points in a transportation network to replace raw and redundant GPS trajectory data points [8]. Popa et al. (2015) proposed an extended data model and a transportation network partitioning algorithm to increase trajectory compression rates without increasing the compression error [9]. The limitation of the first approach is that no optimal solution is provided to adjust the online data sampling rate. Hence, redundant information is captured, transmitted and stored.

The second approach uses a dynamic perspective by reducing the amount of data captured online while not compromising system awareness and control requirements of transportation

This work is supported by the NEXTRANS Center at Purdue University, and CCAT, the Region 5 University Transportation Center.

L. Lin is with the NEXTRANS Center, Purdue University, West Lafayette, IN 47906 USA (e-mail: lin954@purdue.edu).

W. Li is with the Department of Computer Science, University of North Carolina at Chapel Hill, Chapel Hill, NC 27599 USA (e-mail: weizili@cs.unc.edu).

S. Peeta is with the School of Civil and Environmental Engineering, and the H. Milton Stewart School of Industrial and Systems Engineering, Georgia Institute of Technology, Atlanta, GA 30332 USA (e-mail: srinivas.peeta@ce.gatech.edu).

authorities. As an example, the concept of Dynamic Interrogative Data Capture (DIDC) is proposed [10]. The basic idea of DIDC is to identify the lowest data capture and transmission rate while satisfying a certain performance measure request (e.g., system-wide travel time estimation or shockwave location in a specific link). When faced with multiple requests, a DIDC controller will execute a heuristic optimization routine to prioritize the most important one. Though effective, the DIDC controller may cause conflicts among received requests. In addition, the prioritization and sorting of requests are non-trivial. Płaczek [11] developed a framework that dynamically adjusts the data capture rate based on the uncertainty of control decisions instead of performance measures. Data is collected only when the uncertainty is higher than a predefined threshold, and the uncertainty of traffic control decisions is quantified using a fuzzy number comparison approach [12]. The main issue with the second approach is that the data collected is limited to specific tasks and time periods, which entails higher requirements for the scalability and stability of the data analysis algorithms.

In this study, we propose a compressive sensing (CS) based approach for CV data capture and recovery. We also evaluate its impact on travel time estimation. CS has become an active research topic in recent years as a novel approach to capture and recover signals [13]–[15]. Differing from the first approach in which huge amount of data is acquired and compressed, CS enables redundancy removal during the sampling process via a lower but more effective sampling rate [16]. Unlike the second approach, CS does not require dynamic adjustment to the data capture rate based on performance measures [10] or traffic control applications [11]. Instead, it performs a linear transformation to capture the essence of a signal [17], which can then be used to recover the signal for various purposes. The high resemblance of the recovered and original signals allows existing data analysis algorithms to continue functioning without any modifications.

In the transportation domain, CS has been applied only for interpolating missing sensor data from loop detectors or probe vehicles to estimate traffic states. Li et al. (2011) sought to estimate traffic states based on trajectory data of taxis in an urban environment. They applied the CS algorithm for scenarios in which the spatial and temporal trajectory data were missing [18]. Zheng and Su (2016) proposed an algorithm based on CS theory to recover missing traffic flow data from loop detectors and showed that it performs better than a Kalman filter based model [19]. Li et al. (2017) developed a framework based on CS theory to estimate citywide travel times using sparse GPS traces [20]. Our method differs from these studies by applying CS theory for online CV data capture and storage rather than offline processing.

The contributions of this paper are twofold. First, we design a CS-based approach for CV data capture so that less information needs to be stored and transmitted. The proposed approach can be easily implemented with current CV data capture approaches. More specifically, a CV still examines data samples at a fixed rate, except that our approach determines whether to keep a sample or not. Furthermore, the approach allows recovery of the captured data with high accuracy. The proposed approach is evaluated using 10 million CV data samples from the SPMD program. As a result, we can recover the CV data with a root mean squared error (RMSE) of 0.05 by keeping only 20% of the original data (i.e., reducing storage and transmission costs by 80%).

Second, we evaluate the impact of the proposed approach on travel time estimation using a simulation model for a five-mile two-lane freeway segment. In particular, it is compared with two conventional techniques (i.e., using only loop detector data, and using high-sampling CV data) using ground truth values. The proposed approach can generate the most accurate travel time estimations in all simulation scenarios, especially the congested ones. Compared to the high-sampling CV data, the largest relative reduction of travel time estimation error using our approach can reach 65%. Hence, the proposed approach enables a CV to have a smaller on-board unit (OBU) capacity, thus reducing equipment costs.

The rest of the paper is organized as follows. The next section introduces basic CS theory and the proposed CS approach for CV data capture and recovery. Then, two case studies are presented: applying the proposed approach for the compression and recovery of 10 million BSM speed samples, and evaluating the impact of this approach on travel time estimation. The paper concludes with a discussion on the experimental findings and future research directions.

## II. METHODOLOGY

This section introduces the basic concepts of CS theory and the proposed CS approach for data capture and recovery.

### A. Background

Consider a signal vector $x \in R^N$. It can be represented in terms of a set of orthonormal basis $\{\Psi_i\}_{i=1}^{N}$, $\Psi_i \in R^N$ as:

$$x = \Psi\alpha, \qquad (1)$$

where $\Psi$ is an $N \times N$ matrix called Sparsifying Matrix. The signal $x$ is $K$-sparse if $\alpha$, the transformed coefficient vector, has $K$ nonzero entries.

Using the traditional data compression approach (i.e., sample-then-compression), the full signal vector $x$ needs to be acquired first, then the vector $\alpha$ is computed through $\alpha = \Psi^T x$ and only the $K$ largest coefficients are kept [21].

By contrast, CS directly acquires a compressed signal through the following sampling process:

$$y = \Phi x = \Phi\Psi\alpha = \Theta\alpha, \qquad (2)$$

where $\Theta = \Phi\Psi$ is an $M \times N$ matrix. $y \in R^M$ is the sampled vector; $M \ll N$. $\Phi$ is an $M \times N$ matrix called Sensing Matrix.

$M$ and $N$ determine the compression ratio, computed as $M/N$.

Equation (2) defines an underdetermined linear system as the number of equations (i.e., $M$) is much less than the number of unknown entries (i.e., $N$) [16]. The $K$-sparse $x$ can be recovered from $y$ which consists of $M$ measurements by solving the following $l_0$-norm minimization problem:

$$argmin_\alpha \|\alpha\|_0, \text{ subject to } \Theta\alpha = y, \quad (3)$$

where the $l_0$-norm $\|\alpha\|_0$ indicates the number of non-zero elements in the vector denoting the signal's sparsity. Equation (3) is a NP-hard problem and has no efficient solutions.

The CS theory addresses this issue by introducing the following definition [22]: Matrix A satisfies the restricted isometry property (RIP) of order $K$ if there exists a constant $\delta_K \in (0,1)$ such that:

$$(1-\delta_K)\|v\|_2^2 \leq \|Av\|_2^2 \leq (1+\delta_K)\|v\|_2^2, \quad (4)$$

for $\forall v$ satisfying $\|v\|_0 \leq K$.

If the matrix $\Theta$ satisfies the RIP of order $2K$ which can be represented as Equation (5), an accurate reconstruction of a signal can be obtained by solving the following $l_1$-norm optimization problem in Equation (6) [22]:

$$(1-\delta_{2K})\|v\|_2^2 \leq \|\Theta v\|_2^2 \leq (1+\delta_{2K})\|v\|_2^2, \quad (5)$$

where $v = \alpha_1 - \alpha_2$ and $\|v\|_0 \leq 2K$.

$$argmin_\alpha \|\alpha\|_1, \text{ subject to } \Theta\alpha = y. \quad (6)$$

The rationale is that the distance between any pair of $K$-sparse signals $\alpha_1$ and $\alpha_2$ will not be stretched or compressed to a large degree during the dimension reduction from $\alpha \in R^N$ to $y \in R^M$ so that the salient information of a $K$-sparse signal is preserved [21]. Equation (6) can be solved via linear programming or conventional convex optimization algorithms.

*B. CS Approach for CV Data Capture and Recovery*

This section discusses the proposed CS approach for CV data capture and recovery. In particular, it illustrates how to select the matrix $\Psi$ to transform the original CV data vector to a sparse one and the matrix $\Theta$ that satisfies the RIP of order $2K$ to guarantee accurate recovery.

Suppose $x \in R^N$ is a vector of CV data samples, e.g., speed samples collected at a fixed rate. According to Equation (1), we need a transform $\alpha = \Psi^T x$ so that $\alpha$ has a sparse representation in the domain of $\Psi$. Typical transforms include discrete Fourier transform (DFT), discrete cosine transform (DCT), and Discrete Wavelet Transform (DWT). DCT is a Fourier-based transform similar to DFT, but uses cosine functions and the transformed coefficients are real numbers. DWT is more suitable for piecewise constant signals [16], which is not applicable to fluctuating speed samples. Therefore, we select DCT to transform the CV speed signal [23]:

$$\alpha_j = K(j) \sum_{i=1}^N x_i \cos\frac{\pi j(i-0.5)}{N}, j = 0, \ldots, N-1, \quad (7)$$

where $K(j) = \frac{1}{\sqrt{N}}$ when $j = 0$,

$K(j) = \sqrt{\frac{2}{N}}$ when $1 \leq j \leq N-1$.

Next, we need to select a matrix $\Theta$ to obtain the sampled vector $y \in R^M$ (i.e., Equation (2)). As stated earlier, $\Theta$ should satisfy the RIP of order $2K$ so that the original vector $x$ can be recovered. Previous studies have shown the following theorem:

**Theorem 1** [24] Suppose an $M \times N$ matrix $\Theta$ is obtained by selecting $M$ rows independently and uniformly at random from the rows of an $N \times N$ unitary matrix U. By normalizing the columns to have unit $l_2$ norms, $\Theta$ satisfies the RIP with probability $1 - N^{-O(\delta_{2K}^2)}$ for every $\delta_{2K} \in (0,1)$ provided that:

$$M = \Omega(\mu_U^2 K \log^5 N), \quad (8)$$

where $\mu_U = \sqrt{N} \max_{i,j} |u_{i,j}|$ is called the coherence of the unitary matrix U.

Following Theorem 1, we select an $N \times N$ inverse discrete cosine transform (IDCT) matrix $\Psi$ as the unitary matrix U, and randomly select $M$ rows to form the matrix $\Theta$. This allows us to skip the DCT and IDCT transforms and acquire $M$ samples (i.e., $y$) directly from real observations $x$ as follows:

$$y = \Theta\alpha = D\Psi\Psi^T x = Dx, \quad (9)$$

where $\Theta = D\Psi$ represents a random subset of $M$ rows of an $N \times N$ identity matrix.

Finally, after determining matrix $\Psi$ and $\Theta$, the CS approach can be summarized as follows. Suppose a CV is capturing speed samples at a fixed rate. We keep a sample if it is generated from a uniform distribution over $[0,1]$ and is less than or equal to the compression ratio $M/N$. When full data is needed for certain applications, it can be reconstructed by solving the $l_1$-norm optimization problem defined in Equation (6). We illustrate the proposed CS approach for the following two case studies.

### III. CASE STUDY: CAPTURE AND RECOVERY OF 10 MILLION BSM SPEED SAMPLES

The first case study focuses on efficient capture and accurate recovery of 10 million real-world BSM speed samples with the CS approach. The recovery performance of other BSM variables is also evaluated in detail.

*A. Dataset Introduction*

In the SPMD program, BSMs are generated by each CV at 10 Hz [4]. A BSM includes the device ID, timestamp, latitude, longitude, vehicle speed, vehicle heading, yaw rate, and radius of curve. In addition, a BSM includes a "steady state confidence level", which indicates the measurement accuracy; for example, a high confidence level value is commonly found on straight roadways when the CV is in a steady state [25]. In total, we extracted 10 million BSM speed samples. To protect the privacy of the SPMD participants, Personally Identifiable Information (PII) is removed from our dataset. Our dataset consists of 16798 continuous trips, conforming the 10 Hz sampling rate. The mean and standard deviation of the trip speeds are 38.54 *mph* and 24.22 *mph*, respectively.

*B. Sparsity Analysis*

Next, we conduct a sparsity analysis of our dataset. As an illustration, Figs. 1(a) and 1(b) show a set of BSM speed samples $x$ with $N = 1000$ and the corresponding DCT coefficients $\alpha$. Only 157 out of 1000 coefficients are greater than 1, while the others are negligible. This indicates $\alpha$ is indeed sparse, implying we can apply the CS approach to capture CV data via a lower sampling rate.

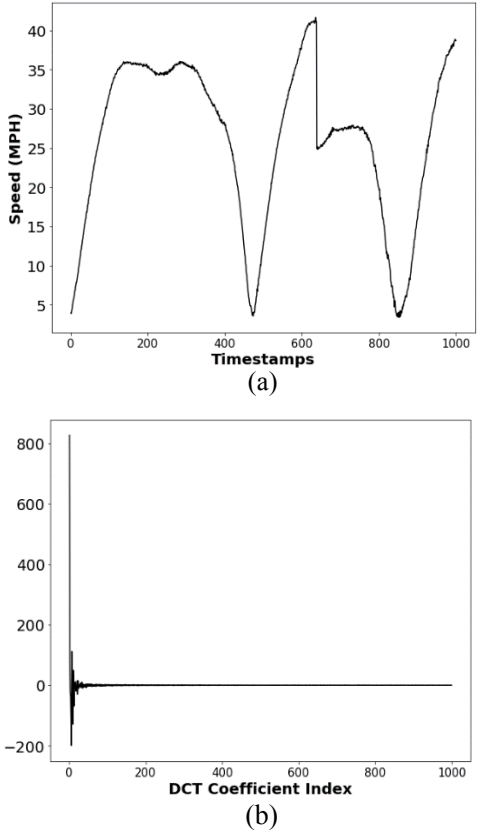

Fig. 1. (a) the original 1000 BSM speed samples; (b) the 1000 DCT coefficients.

### C. Recovery Accuracy Evaluation Criterion

When the full data is needed for an application such as travel time estimation, it can be recovered by solving Equation (6) to convert $y \in R^M$ to coefficients in DCT domain $\alpha \in R^N$. The IDCT is then performed on $\alpha$ to obtain the recovered data $\hat{x} \in R^N$. The recovery process can be easily implemented in parallel. The recovery accuracy is measured by calculating the root mean squared error (RMSE) normalized with respect to the $l_2$-norm of the entire data series [16]:

$$RMSE = \frac{\|x_o - \hat{x}_r\|_2}{\|x_o\|_2}, \qquad (10)$$

where $x_o$ is the original 10 million speed samples and $\hat{x}_r$ is the recovered BSM speed data.

### D. Recovery Performance related to M and N

The key parameters of our approach are $M$ and $N$, which determine the compression ratio and affect the recovery accuracy. Fig. 2(a) shows the RMSEs of the 10 million speed samples under different compression ratios.

When the compression ratio is 0.1, the RMSE of $N = 1000$ is the lowest. As the compression ratio increases, the RMSEs become lower and close to each other for different values of $N$. When the compression ratio is greater than 0.2, the RMSEs are capped at 0.025. When the compression ratio reaches 0.6, all RMSEs are close to zero.

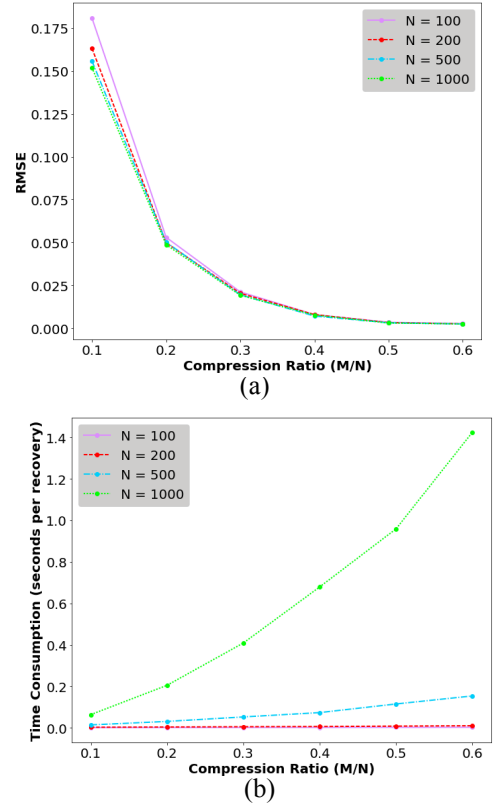

Fig. 2. (a) RMSE by compression ratio $(M/N)$; (b) Time per recovery by compression ratio $(M/N)$.

The computational complexity of $l_1$-norm optimization in Equation (6) is $O(N^3 + MN^2)$ [21]. The average time per recovery is also calculated and shown in Fig. 2(b). All experiments are conducted in Windows 10, using i7-6820HK CPU and 64 GB RAM.

The time per recovery is close to zero for all compression ratios when $N = 100$ and $N = 200$. When $N = 500$ and $N = 1000$, the curves of time per recovery have a big increase under larger compression ratios. In particular, the time is higher than 1.4 seconds per recovery when the compression ratio is equal to 0.6 and $N = 1000$.

Based on the above analysis of RMSEs and computational efficiency, we select $M = 40$ and $N = 200$ for this case study. The corresponding RMSE calculated using Equation (10) under these parameter values is about 0.05.

To illustrate the effect of our approach on CV data collection, a trip made by CV number "2300" is selected. The trip originally has 4967 speed samples. After applying the compression ratio 0.2 (M=40 and N=200), only 993 samples are retained. Fig. 3(a) shows locations of some original speed samples (marked in black) and Fig. 3(b) shows locations of the samples distilled using our approach (marked in red). Fig. 3(c) further shows the original 4967 speed samples (black) and the corresponding recovered samples (red). The recovered data highly resemble the original data with only 0.02 RMSE.

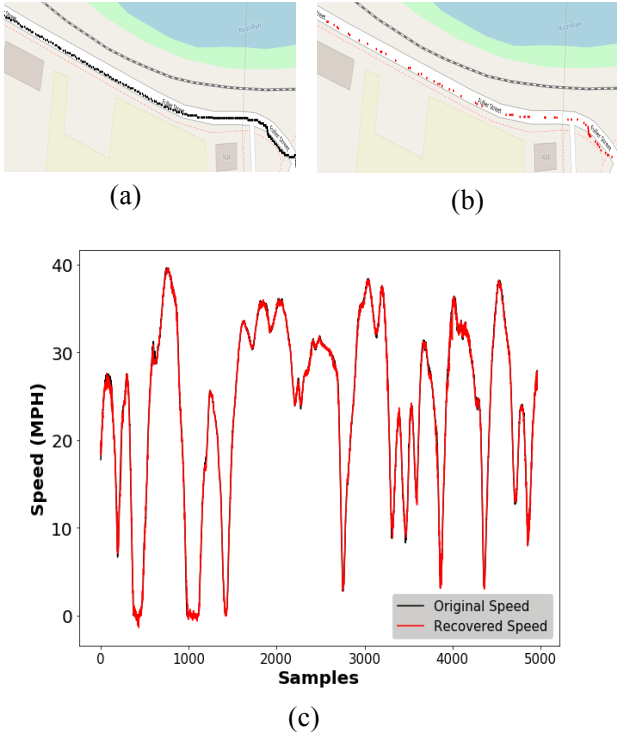

Fig. 3. (a) Locations of original speed samples (black), (b) Locations of compressed speed samples from the CS approach (red), and (c) Original and recovered speed samples.

*E. Recovery Performance of Other BSM Variables*

We further explore the recovery performance of the CS approach for other BSM variables, e.g., various speed categories. The original 10 million speed samples are split into 8 categories by every 10 MPH and the corresponding RMSE and the number of samples in each category are calculated. Fig. 4(a) shows that most speed categories have more than 1 million samples except the "11-20" and "51-60" categories. The RMSE curve decreases substantially as speed category increases. This implies the CS approach performs better in high speed situations. Note that the actual traffic states of the original data are unknown. So, it may not be reasonable to apply speed category as a proxy for actual traffic states.

The recovery performance of the CS approach is evaluated in terms of driving conditions categorized by yaw rate. In total, there are 12 yaw rate categories from [−360, 360] at 60 degrees interval. Fig. 4(b) shows the RMSEs and number of samples in each yaw rate category. A negative yaw rate implies that a CV is turning to the left while a positive value indicates it is turning to the right [25]. Most yaw rates fall into [-60, 0) and [0, 60) categories. For other categories, only few thousand samples exist. The RMSEs are close to zero for categories [-60, 0) and [0, 60) and become larger when yaw rate is higher.

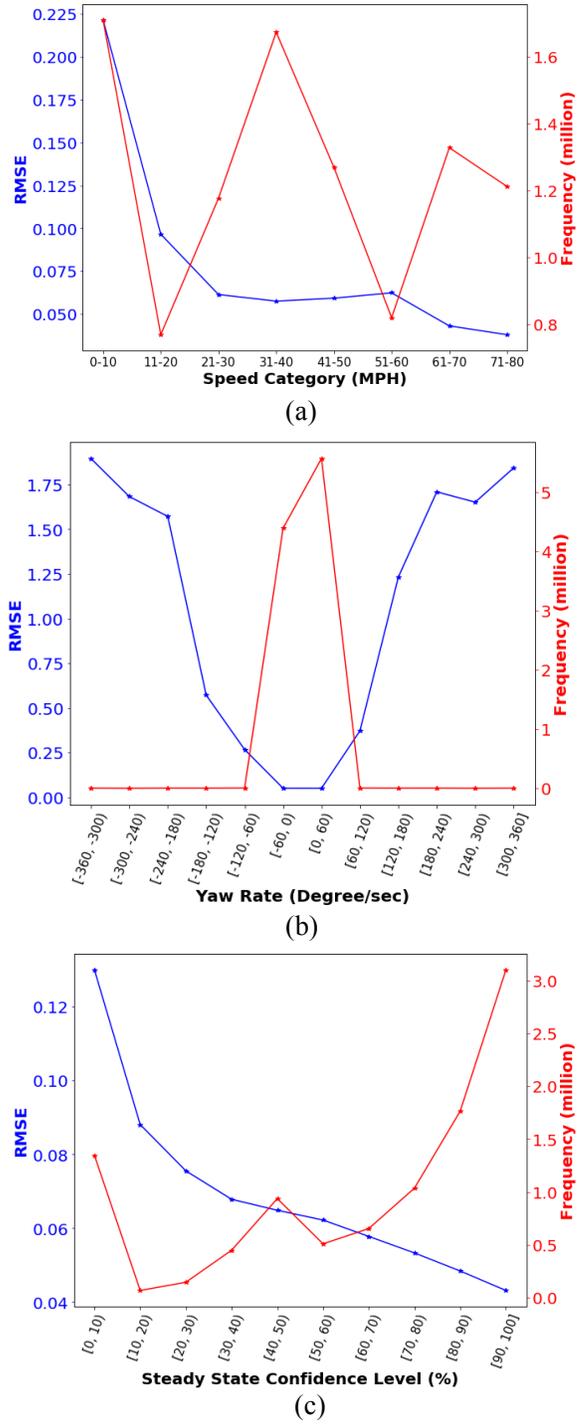

Fig. 4. (a) Recovery performance by speed category, (b) Recovery performance by yaw rate category, and (c) Recovery performance by steady state confidence level category.

Further, the correlation between the recovery accuracy using the CS approach and the BSM variable "steady state confidence level" is examined. The confidence level interval [0,100] is uniformly divided into 10 categories. Fig. 4(c) shows the RMSEs and number of samples for each steady state confidence level category. There are more samples in higher steady state confidence level categories: about 5 million speed samples have more than 80% confidence level. The RMSE curve decreases when the confidence level increases. This shows the quality of the CV data impacts the recovery performance. Data with higher measurement error cannot be recovered as accurately as that with minor measurement error.

In this case study, the proposed CS approach is evaluated using 10 million BSM speed data from the SPMD program. However, it is difficult to evaluate the impact of this approach on a certain application (such as travel time estimation) without ground truth traffic data. To do so, in the next case study, we build a traffic simulation model.

## IV. CASE STUDY: IMPACT OF THE CS APPROACH ON TRAVEL TIME ESTIMATION

This case study evaluates the impact of the proposed CS approach on travel time estimation through a traffic simulation model. Travel time estimation aims to provide the travel time from one point to another in a link for a certain time interval [26]. Accurate and reliable travel time estimation plays a critical role in active traffic management [27].

### A. Travel Time Estimation from Various Data Sources

Using a simulation model, for each segment $s$ and each time interval $j$, we can estimate travel times from three data sources: traditional loop detector data, CV data captured at a fixed rate, and CV data via the CS approach. For convenience, these travel times are referred to as $TT_{LP}^{s,j}$, $TT_{CV}^{s,j}$ and $TT_{CS}^{s,j}$ hereafter. As the trajectory data of each vehicle is also available, the ground truth travel times, denoted as $TT_{GR}^{s,j}$, can also be acquired. The computation of these quantities is as follows:

$$TT_d^{s,j} = \frac{L_s}{\bar{v}_d^{s,j}}, \tag{11}$$

where $d = GR, LP, CV$ or $CS$, indicating the data sources, $L_s$ is the length of segment $s$, and $\bar{v}_d^{s,j}$ is the space mean speed over the segment $s$ at time interval $j$ based on data source $d$.

The space mean speed $\bar{v}_d^{s,j}$ is calculated in two ways in the literature. Double-loop detector is a key source to measure vehicle speed [28]. In this case, $\bar{v}_d^{s,j}$ is defined as the harmonic mean of the speeds of vehicles passing the loop detector at the end of segment $s$ during a time interval $j$, which is calculated as [29].

$$\bar{v}_d^{s,j} = \frac{N_{s,j}}{\sum_{i=1}^{N_{s,j}} \frac{1}{v_{i,j}}}, \tag{12}$$

where $d = LP$, $N_{s,j}$ is the number of vehicles passing the loop detector at the end of segment $s$ at time interval $j$, and $v_{i,j}$ is the speed of vehicle $i$ passing the loop detector at the end of segment $s$ at time interval $j$.

When the trajectory data from probe vehicles such as CVs is available, some studies define the space mean speed as the average of the mean speeds of all probe vehicles over segment $s$ at an instant of time $t$ within a time interval $j$ [16], [18]:

$$\bar{v}_d^{s,j} = \frac{\sum_{t=1}^{T_j}\left(\frac{\sum_{i=1}^{N_{s,t}} v_{i,s,t}}{N_{s,t}}\right)}{T_j}, \tag{13}$$

where $d = GR, CV$ or $CS$ represents the data source for providing the data, $T_j$ is the number of time steps in time interval $j$, $N_{s,t}$ is the number of probe vehicles at segment $s$ at time step $t$, and $v_{i,s,t}$ is the speed of probe vehicle $i$ at segment $s$ at time step $t$.

### B. Traffic Simulation Model Setup

To evaluate the impact of the CS approach on travel time estimation, a simulation model is built for a five-mile two-lane freeway segment using SUMO. SUMO is an open-source microscopic simulator, which provides rich inter-vehicle interactions [31]. The layout of the freeway segment is shown in Fig. 5.

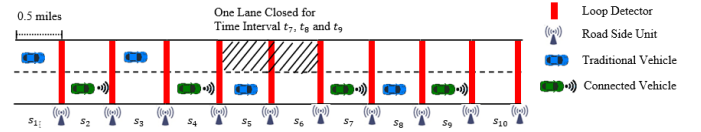

Fig. 5. Five-mile two-lane freeway segment.

The five-mile freeway segment consists of 10 small segments, $s_1$ to $s_{10}$, of length 0.5 miles each. One loop detector and one RSU are located at every 0.5 mile. The total simulation time is 3600 seconds, and the traffic demand is set at 2400 vehicles. The traditional, non-connected vehicles (blue) and CVs (green) are generated based on a predefined CV penetration rate. The simulation period is split into 12 time intervals of 300 seconds, $t_1, \ldots, t_{12}$. The first 3 time intervals are considered as the warm-up period, and thus excluded from the analysis. In time intervals $t_7$, $t_8$ and $t_9$, we close the inner lanes of segments $s_5$ and $s_6$ to create congestion conditions. Additional assumptions are as follows:

1. CV On-board Units (OBUs) are assumed to have a limited capacity. According to Kianfar and Edara (2013), the OBU can store up to 30 snapshots including the vehicle speed, location, and the time stamp according to the SAE J2735 standard are stored [30]. In this case study, we conduct analysis under different OBU snapshot capacities.

2. If the OBU capacity is reached before the CV passes a road-side unit (RSU), earlier recordings will be replaced by later snapshots, causing information loss [32].

3. When a CV passes a RSU, all recorded snapshots are transmitted to the RSU instantly. A RSU can communicate with multiple CVs at the same time [32]. No transmission loss or delay are considered.

4. All loop detectors and RSUs used in the simulation model operate normally. The loop detectors can record accurate vehicle speeds.

5. After the CV data captured using our approach is transmitted to RSUs and uploaded to the transportation management center, the recovery operation is executed. Travel times are then estimated using the recovered CV data.

*C. Travel Time Estimation Accuracy Evaluation Criterion*

Using Equations (12) - (14), the overall Mean Absolute Percentage Errors (MAPE) of travel time estimation $TT_{LP}$, $TT_{CV}$ or $TT_{CS}$ is calculated as follows:

$$MAPE_d = \frac{\sum_{s=2}^{N}\sum_{j=4}^{T}\frac{|TT_d^{s,j}-TT_{GR}^{s,j}|}{TT_{GR}^{s,j}}}{(N-1)*(T-3)}, \quad (14)$$

where $d = LP, CV$ or $CS$ indicates the three data sources for travel time estimation, $N = 10$ because there are 10 small segments of length 0.5 miles each, $T = 12$ because the simulation time is split into 12 time intervals of 300 seconds each, $i$ starts from 2 because vehicles enter the segment $s_1$ with a speed of zero in SUMO, and $j$ starts from 4 because the first 3 time intervals are considered as warm-up period.

*D. Sensitivity Analysis*

Various parameters in the simulation model can be adjusted: OBU Capacity, Compression Ratio, CV Data Capture Rate, CV MPR and Arrival Rate.

We first set the CV MPR at a fixed value 0.6 and define OBU capacity set as {50, 100, 150, 200, 250, 300} snapshots, CV data capture rate set as {1, 10} Hz, and Compression Ratio set as {0.2, 0.5}. For the compression ratio 0.2, $M$ and $N$ are set as 40 and 200, and for the compression ratio 0.5, $M$ and $N$ are set to 100 and 200, respectively. Two Arrival Rate patterns are tested. One uses a fixed arrival rate 2400 vehicles/hour, and the other uses a varying arrival rate: 1200 vehicles/hour for $t_1$-$t_3$, 2400 vehicles/hour for $t_4$-$t_6$, 4800 vehicles/hour for $t_7$-$t_9$, and 1200 vehicles/hour for $t_{10}$-$t_{12}$ (the total demand is still 2400 vehicles). Fig. 6 shows the performances of $TT_{LP}$, $TT_{CV}$, and $TT_{CS}$ in different scenarios when the CV MPR is 60%. For each scenario, we run the simulation 5 times and compute the average $MAPE_{LP}$, $MAPE_{CV}$ and $MAPE_{CS}$.

First, in most scenarios, $MAPE_{CV}$ and $MAPE_{CS}$ are lower than $MAPE_{LP}$, except the few cases when the data capture rate is 10 Hz and the OBU capacity is set to either 50 or 100 snapshots as shown in Fig. 6(b). This is mainly because only limited road information can be stored in a CV at a high data capture rate using a small OBU capacity. One exception is that when CVs collect data via our approach with a compression ratio 0.2, although the OBU capacity is as small as 50, we observe a lower MAPE than $TT_{LP}$.

Second, as shown in all subplots of Fig. 6, the $MAPE_{CV}$ and $MAPE_{CS}$ curves decrease as OBU capacity increases. Using the same simulation setting, $MAPE_{CS}$ is always lower than $MAPE_{CV}$. This is because our approach offers broader spatio-temporal coverage and can accurately recover the CV data. The simulation results verify that trading a little accuracy for broader spatial-temporal coverage is beneficial for travel time estimations using CV data.

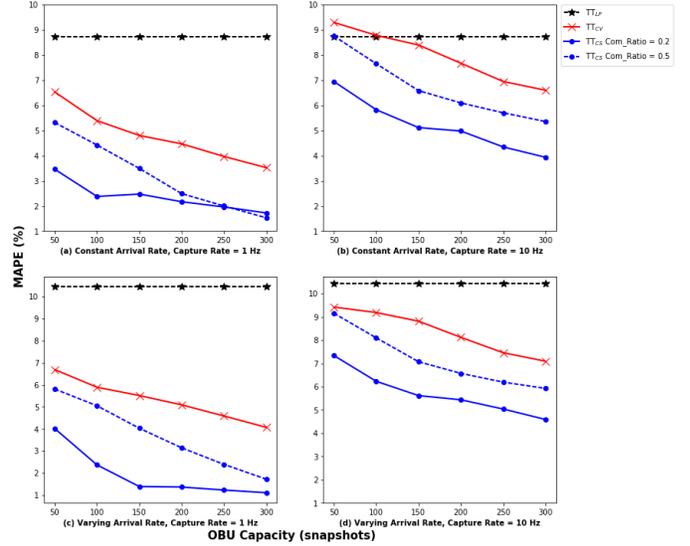

Fig. 6. Travel time estimation performance by OBU Capacity, Arrival Rate, Compression Ratio, and Data Capture Rate (CV MPR = 60%).

Third, by comparing Fig. 6(b) to Fig. 6(a), when the data capture rate changes from 10 Hz to 1 Hz, the MAPE performances of $TT_{CV}$ and $TT_{CS}$ are improved. This indicates a higher CV data capture rate (e.g., 10 Hz) may not be necessary for travel time estimation when limited OBU capacity is available. The same observation can be made by comparing Figs. 6(d) and 6(c).

Third, when the data capture rate is 1 Hz, comparing the $MAPE_{CS}$ curves in Fig. 6(a) indicates that for all the OBU capacities except "300", the one with the compression ratio 0.2 performs better than the one with the compression ratio 0.5. Once again, this is due to the broader spatio-temporal coverage provided by the proposed CS approach. Similarly, in Fig. 6(c) when the arrival rate varies, the $TT_{CS}$ with a compression ratio 0.2 always performs better than the one with a compression ratio 0.5. This suggests that a smaller compression ratio does not require the CV to have a large OBU capacity to reach the same travel time estimation accuracy. Therefore, the hardware costs of OBUs can be reduced using our technique.

Next, we fix the CV Data Capture Rate at 1 Hz and explore the impact of OBU Capacity, Arrival Rate, Compression Ratio, and CV MPR on travel time estimation. First, the decreasing pattern is not as dramatic as the increasing pattern of CV MPR, irrespective of the values of other quantities. Second, $MAPE_{CS}$ is consistently lower than $MAPE_{CV}$ in all scenarios, demonstrating the effectiveness of our approach. Third, in Figs. 7(a) and 7(c) when the OBU Capacity is 50, $TT_{CS}$ with compression ratio 0.2 is more accurate than the value with compression ratio 0.5. By contrast, in Figs. 7(b) and 7(d) when the OBU capacity is 300, the two $MAPE_{CS}$ curves are closer to each other. This indicates that a lower compression ratio would allow a CV with a small OBU Capacity to cover a larger road segment and for a longer period of time. Nevertheless, when the OBU Capacity is large enough, the effect of the compression ratio is not as prominent.

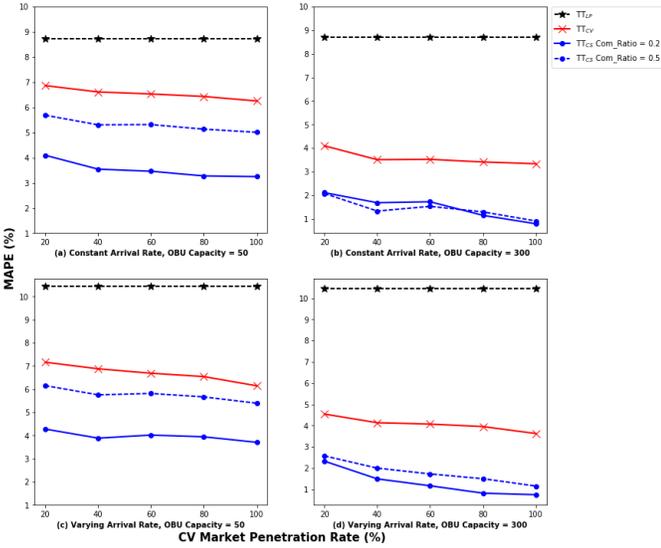

Fig. 7. Travel time estimation performances by OBU Capacity, Arrival Rate, Compression Ratio, and CV MPR (CV Data Capture Rate = 1 Hz).

To further analyze the performance of the proposed CS approach, we fix the CV Data Capture Rate, Compression Ratio, and OBU Capacity and test all combinations of Arrival Rate and CV MPR by running each scenario 5 times. Fig. 8(a) shows the means and standard deviations of $MAPE_{CV}$ and $MAPE_{CS}$ based on the parameter "CV Data Capture Rate-Compression Ratio-OBU Capacity". The means of $MAPE_{CV}$ and $MAPE_{CS}$ decrease as the Data Capture Rate decreases from 10 Hz to 1 Hz, the Compression Ratio decreases from 0.5 to 0.2, and the OBU Capacity increases from 50 to 300 snapshots. The mean of $MAPE_{CS}$ is consistently lower than the mean of $MAPE_{CV}$. Except for the scenario of "10-0.5-50", even the upper bound of $MAPE_{CS}$ is lower than the lower bound of $MAPE_{CV}$. Fig. 8(b) further calculates the relative reduction by comparing the mean of $MAPE_{CS}$ to the mean of $MAPE_{CV}$. The result shows that when Data Capture Rate is 10 Hz, the relative reduction is always lower than 40%. Generally, the relative reduction is higher when the Data Capture Rate is 1 Hz; in particular, the relative reduction ranges from 43% to 65% when the "Data Capture Rate-Compression Ratio-OBU Capacity" changes from "1-0.2-50" to "1-0.2-300".

### E. Impact of Lane Closing

As stated earlier, to create congestion conditions, the inner lanes in $s_5$ and $s_6$ are closed during time intervals $t_7$, $t_8$, and $t_9$. Fig. 9 shows the map of the ground truth vehicle speed by segment and time interval from simulation results with constant arrival rates. Starting from time interval $t_7$, a shockwave is observed moving backwards from upstream segment $s_4$ to $s_2$, where the traffic state changes from a free-flow state to a congested state.

To evaluate the impact of the traffic state transition on travel time estimation, the $MAPE_{CS}$ and $MAPE_{CV}$ are computed by segment and time interval when CV MPR = 60% and OBU capacity = 300. Fig. 10 illustrates that due to congestion, travel time estimation has the least accuracy on the upstream segments $s_2$-$s_4$ from the 8th time interval. In both Figs. 10(a) and 10(b), the highest MAPE occurs on Segment 3 in the 9th time interval. A comparison of Figs. 10(a) 10(b) shows that, in most cases, travel times $TT_{CS}$ are more accurate than $TT_{CV}$ on the upstream segments starting from the 8th time interval. This is because if a CV with limited OBU capacity is collecting data at a fixed rate and is stuck in a traffic jam, the OBU will be saturated with close to 0 speed readings. By comparison, the proposed CS approach offers broader spatio-temporal coverage, and hence provides more accurate results reflected by the lower MAPE.

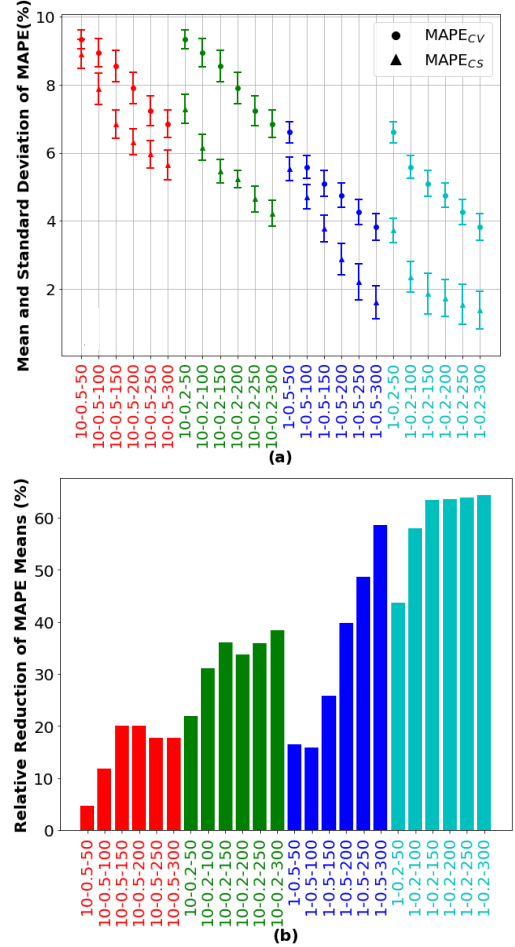

Fig. 8. (a) Mean and standard deviation of $MAPE_{CV}$ and $MAPE_{CS}$ by CV Data Capture Rate-Compression Ratio-OBU Capacity, and (b) Relative reduction of comparing means of $MAPE_{CS}$ and $MAPE_{CV}$ by CV Data Capture Rate-Compression Ratio-OBU Capacity.

## V. CONCLUSION AND FUTURE RESEARCH DIRECTIONS

As connected vehicles (CVs) become more widespread, huge amounts of data are being collected, stored, and transmitted. Hence, there exists the possibility of redundant information that overwhelms storage and communications, resulting in prohibitive costs. To mitigate this critical issue, we propose a Compressive Sensing (CS) based approach for CV data collection and recovery. Our technique allows CVs to compress data in real-time, and can accurately and efficiently recover the original data. The propose approach is evaluated using two comprehensive case studies, which demonstrate its effectiveness and efficiency, and use in applications such as travel time estimation.

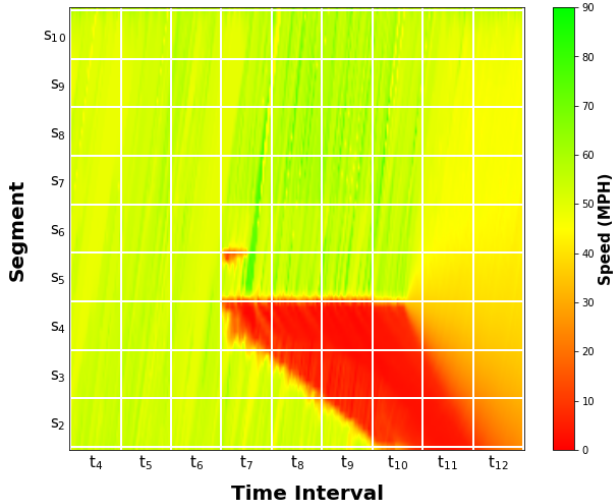

Fig. 9. Map of ground truth vehicle speed by segment and time interval.

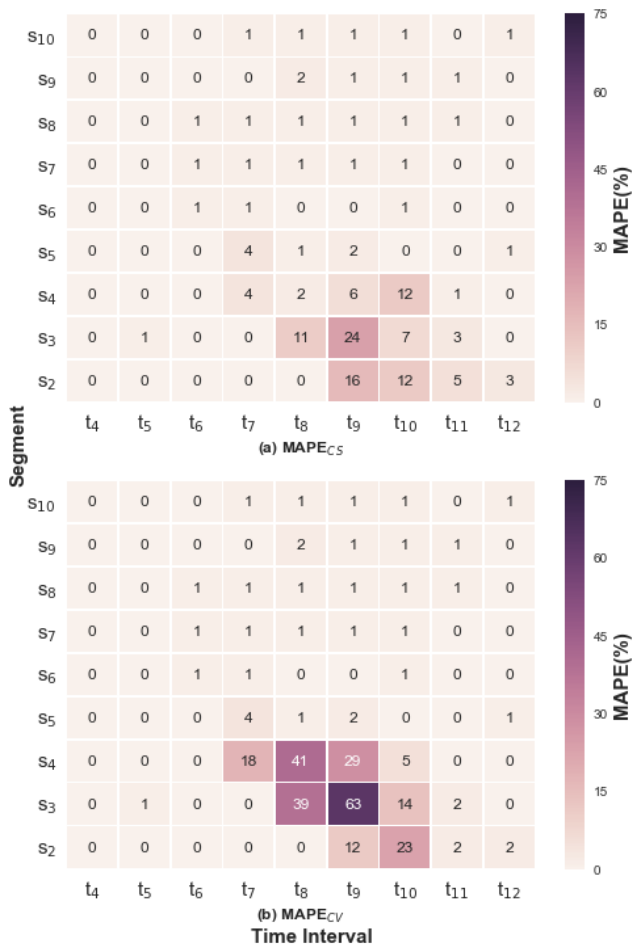

Fig. 10. MAPEs by segment and time interval (CV MPR = 60%, OBU Capacity = 300, Data Capture Rate = 1 Hz).

In the first case study, 10 million Basic Safety Message speed samples are extracted from the SPMD program. When the compression ratio is 0.2 ($M = 40$ and $N = 200$), the CS approach can efficiently recover the original speed data with RMSE as low as 0.05. The recovery performance of the approach is also evaluated for speed, yaw rate, and steady state confidence level. The results suggest that the approach performs better when the CV is driving with a high speed or a low yaw rate. The approach also performs better with a high steady state confidence level, which indicates that measurement error is small.

The second case study uses a five-mile two-lane freeway simulation model to evaluate the impact the approach on travel time estimation. Congestion is created by shutting down one lane in the middle of the freeway for certain periods of time. Multiple scenarios under various CV Market Penetration Rates, On-board Unit Capacities, Compression Ratios, Arrival Rates, and Data Capture Rates are simulated. In each scenario, the travel times are estimated from different sources: traditional loop detector data, CV data collected with a fixed rate, and CV data using our approach. The proposed CS approach consistently obtains the lowest MAPE when compared to the ground truth values. For certain scenarios, the relative MAPE reduction of applying the CS approach instead of conventional data collection methods can reach 43% to 65%. The study also shows that this approach offers broader spatio-temporal coverage when used with a small compression ratio. In addition, it can recover travel times accurately using a small OBU capacity and scales well when the OBU capacity increases. This implies that the OBU costs can be reduced using the CS approach. Also, it significantly improves the estimation accuracy in a congested environment.

There are several possible future research directions. First, the impact of the proposed approach can be analyzed for other CV applications such as information propagation through real-time V2V communications. Second, the CS approach can be applied for multimodal data capture in autonomous vehicles which have multiple sensors such as cameras and LIDAR. Third, network simulator NS-3 can be incorporated to build more realistic simulation scenarios. Fourth, the impact of factors such as transmission loss and delay on the CS approach can be examined. Finally, it would be interesting to derive a dynamic compression ratio ($M/N$) based on data quality and driving conditions (such as low steady state and speed).

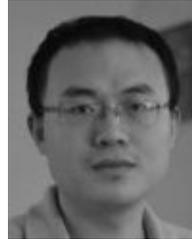

**Lei Lin** received the B.S. degree in traffic and transportation and the M.S. degree in system engineering from Beijing Jiaotong University, China, in 2008 and 2010, respectively. He received the M.S. degree in computer science and the Ph.D. degree in transportation systems engineering from the University at Buffalo, the State University of New York, Buffalo, in 2013 and 2015, respectively. From 2013 to 2015, he was a Research Assistant with Transportation Informatics Tier 1 University Transportation Center. He worked as a researcher for Xerox from 2015 to 2017. Since 2017, he has been a research associate at the NEXTRANS Center, Purdue University. His research interests include transportation big data, machine learning applications in transportation, and connected and automated transportation.

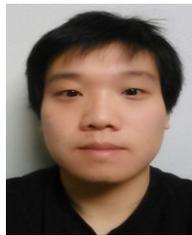

Weizi Li received his B.E. degree in Computer Science and Technology from Xiangtan University, China and M.S. degree in Computer Science from George Mason University. He is currently in the doctoral program at the University of North Carolina at Chapel Hill, Department of Computer Science. His research interests include agent-based simulation, intelligent transportation systems, and statistical machine learning.

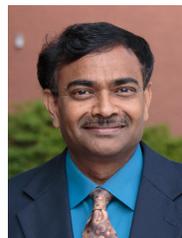

**Srinivas Peeta** received his M.S. and Ph.D. degrees in civil engineering from Caltech and the University of Texas at Austin, U.S.A., respectively, in 1989 and 1994. He is the Frederick R. Dickerson Chair and Professor in the Schools of Civil and Environmental Engineering and Industrial and Systems Engineering at Georgia Institute of Technology. He is also Principal Research Faculty at the Georgia Tech Research Institute. Prior to that, he was the Jack and Kay Hockema Professor in Civil Engineering at Purdue University and the Director of the NEXTRANS Center, formerly the U.S. Department of Transportation's (USDOT's) Federal Region 5 University Transportation Center. He also served as the Associate Director of USDOT's Center for Connected and Automated Transportation at Purdue University. His research interests include intelligent transportation systems (ITS), operations research, control theory, and computational intelligence techniques. He serves on the Editorial Advisory Boards of the journals Transportation Research Part B, and Intelligent Transportation Systems Journal. He serves as area editor for Transportation Dynamics for Networks and Spatial Economics. He has previously served as Chair of the Transportation Network Modeling Committee of the Transportation Research Board of the National Academies. He is also a member of International Federation of Automatic Control Technical Committee on Transportation Systems.